\documentclass[acmsmall]{acmart}
\usepackage{xcolor}

\title{Making Embodied AI Reliable: A Community Agenda from Testing to Formal Verification}

\author{Xi Zheng,
Dulanga Weerakoon,
Yintong Huo,
Teresa Yeo,
Guy Van den Broeck,
Vijay Ganesh,
Daniel Neider,
Biplav Srivastava,
Ivan Ruchkin,
Archan Misra,
Corina Pasareanu
}

\begin{document}

\begin{abstract}
Embodied AI systems are increasingly deployed in open-world environments, yet ensuring their reliability remains a fundamental challenge. Drawing on discussions from the AAAI'26 Bridge Program on ``Making Embodied AI Reliable with Testing and Formal Verification'', this article argues that reliability in embodied AI is inherently a lifecycle assurance problem arising from uncertainty, human interaction, and emergent behaviors across tightly coupled system components. We identify three complementary directions toward reliable embodied AI: (i) trustworthy scenario-based testing grounded in validated specifications and coverage-driven evaluation, (ii) compositional verification enabled through structured symbolic abstractions, and (iii) runtime assurance mechanisms capable of adapting under uncertainty and distribution shifts. Rather than treating these approaches independently, we advocate for integrated assurance workflows that connect testing, verification, and runtime adaptation through shared neuro-symbolic representations and closed-loop feedback across the system lifecycle. Such integration provides a foundation for trustworthy embodied AI systems operating safely in complex real-world environments.
\end{abstract}

\maketitle

\section{Reliability as the Central Challenge in Embodied AI}
Discussions at the AAAI'26 Bridge Program on ``Making Embodied AI Reliable with Testing and Formal Verification'' highlighted four fundamental challenges for embodied AI reliability: uncertainty caused by imperfect perception, human interaction involving unpredictable human behaviors, dynamic environments where open-world conditions can lead to machine-learning component failures and random errors, and emergent behaviors caused by cross-module interactions among perception, reasoning, planning, and control. These challenges expose important limitations of existing assurance methodologies when applied to embodied AI systems operating in complex real-world environments.

\begin{figure}[t]

    \centering

    \includegraphics[width=\linewidth]{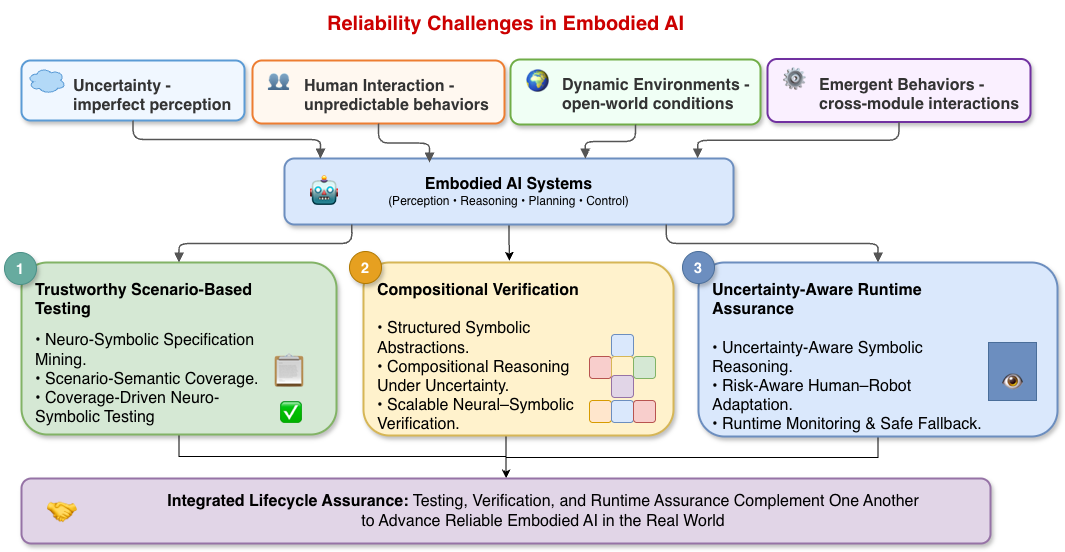}

    \caption{Overview of key reliability challenges in embodied AI and the three complementary research directions.}

    \label{fig:overview}

\end{figure}

Figure~\ref{fig:overview} summarizes this perspective and highlights three complementary directions toward reliable embodied AI: trustworthy scenario-based testing grounded in validated specifications and coverage-driven evaluation, compositional verification through structured symbolic abstractions, and uncertainty-aware runtime assurance. Rather than treating these approaches independently, reliable embodied intelligence requires integrated lifecycle assurance workflows spanning specification mining, testing, verification, deployment monitoring, and adaptive decision-making.

\section{Scenario-Based Testing and LLM-Guided Specification Mining}

Embodied AI systems operate under uncertainty, interact with humans, and continuously adapt to dynamic open-world environments. These characteristics make scenario-based testing a primary assurance mechanism for evaluating safety-critical behaviors in domains such as autonomous driving and robotics. Unlike traditional software systems, embodied AI testing must reason about interactions among perception, reasoning, planning, environmental dynamics, and human behavior, making the quality and semantic validity of test scenarios fundamentally important.

Large language models (LLMs) provide a promising pathway for translating natural-language requirements into structured scenarios, but purely neural generation pipelines remain prone to hallucination, inconsistency, and prompt sensitivity~\cite{zheng2024testing, cosler2023nl2spec}. Unrealistic object relationships, infeasible environmental constraints, and incomplete interaction patterns can produce misleading or invalid scenarios, limiting testing reliability and often requiring human validation.

To address these limitations, emerging approaches increasingly incorporate symbolic constraints and structured world representations directly into the specification mining process. Rather than relying on unconstrained prompting, \emph{neuro-symbolic specification mining} integrates domain rules, scene graphs, physical properties, interaction constraints, and domain-specific languages (DSLs) into scenario generation~\cite{li2024guiding}. These structured representations capture not only semantic object relationships but also interaction dynamics, environmental constraints, and uncertainty-aware contextual information. In turn, this improves semantic consistency, reduces hallucination, and provides a common interface for scenario generation, validation, and downstream analysis.


A related challenge concerns how to define meaningful testing coverage for embodied AI systems. Traditional metrics such as neuron coverage often lack interpretability and show weak correlation with defect discovery~\cite{hildebrandt2023physcov}. Increasingly, attention is shifting toward \emph{scenario-semantic coverage}, where coverage is defined over objects, spatial relations, interactions, trajectories, and environmental conditions. The core challenge is therefore no longer generating large numbers of scenarios, but systematically exploring semantically structured regions of scenario space associated with safety-critical behaviors.

These challenges motivate \emph{coverage-driven neuro-symbolic testing} pipelines, where LLM-generated scenarios are refined using symbolic constraints, coverage objectives, uncertainty estimation, and failure feedback~\cite{he2024curiosity}. Automated oracles combining symbolic rules, physical constraints, and learned models can further validate generated scenarios and execution traces. More broadly, such structured neuro-symbolic representations provide a common semantic foundation across testing, verification, and runtime assurance, enabling embodied AI assurance to evolve from isolated validation techniques toward integrated lifecycle assurance workflows.

\section{Compositional Verification and Modular Reasoning}

Formal verification of embodied AI systems remains fundamentally constrained by the lack of semantically meaningful verification abstractions. High-level goals such as ``do not harm humans'' are often too ambiguous to formalize precisely, while low-level representations such as individual pixels, raw sensor signals, or neural activations lack semantic interpretability. As a result, existing verification techniques struggle to bridge neural perception, symbolic reasoning, and system-level behavioral guarantees in dynamic open-world environments.

A promising direction is the use of \emph{structured symbolic abstractions} as semantic interfaces between neural and symbolic components. Rather than reasoning directly over raw neural representations, embodied AI systems can construct structured world representations---such as scene graphs, object relations, trajectories, and environmental constraints---that capture semantically meaningful system and environmental states. These abstractions provide interpretable interfaces between perception, reasoning, planning, and control modules while enabling downstream formal reasoning over symbolic structures rather than uninterpretable neural activations.

Additionally, embodied AI verification must support \emph{compositional reasoning under uncertainty}. Perception outputs are inherently uncertain and incomplete, while environmental dynamics and distribution shifts may invalidate assumptions embedded within downstream reasoning or planning modules. Consequently, symbolic abstractions must represent not only semantic structure, but also uncertainty, ambiguity, and incomplete world knowledge. Emerging approaches increasingly combine neural robustness analysis with probabilistic symbolic world models, enabling uncertainty-aware reasoning across heterogeneous neural and symbolic pipelines.

Building on these abstractions, \emph{scalable neuro-symbolic verification} requires compositional reasoning mechanisms capable of propagating assumptions, uncertainties, guarantees, and operational objectives across interacting system components. Rather than attempting computationally intractable end-to-end verification, embodied AI systems can instead leverage modular and assume-guarantee reasoning techniques to analyze how local guarantees and competing objectives interact across perception, reasoning, planning, and control modules. Such approaches would provide a more scalable and interpretable foundation for verifying complex embodied systems, whose failures often emerge not only from isolated component behaviors, but also from inappropriate prioritization and coordination across interacting modules.

More broadly, embodied AI assurance requires moving beyond static module verification toward integrated reasoning about uncertainty propagation, inter-module dependencies, and evolving environmental interactions. Therefore, structured symbolic abstractions would serve not merely as auxiliary representations, but as foundational semantic interfaces connecting neural perception, formal reasoning, runtime monitoring, and system-level assurance across the embodied AI lifecycle.

\section{Runtime Assurance Under Uncertainty and Human-Robot Interaction}

Embodied AI systems operate in dynamic open-world environments where uncertainty, environmental change, and unpredictable human behavior continuously challenge system reliability. While scenario-based testing and compositional verification can provide partial guarantees under bounded assumptions, real-world deployment may violate these assumptions through severe distribution shifts, sensor degradation, unforeseen environments, or evolving human interactions. As a result, runtime assurance becomes essential for maintaining reliability beyond the conditions explicitly considered during design-time validation.

A central capability for runtime assurance is \emph{uncertainty-aware symbolic reasoning}. Learning-based systems provide scalability and generalization capabilities, but often rely on statistical correlations that may fail under changing environments. Instead, neuro-symbolic approaches combine learning-based perception with structured symbolic world models that represent objects, relations, environmental constraints, causal dependencies, and system assumptions~\cite{khan2025survey}. Unlike purely deterministic symbolic reasoning, uncertainty-aware symbolic reasoning enables embodied systems to reason over ambiguous, incomplete, or probabilistic world states during deployment~\cite{rating-composition-planning}. Uncertainty originating from perception, localization, prediction, or human interaction can therefore propagate into symbolic reasoning and downstream decision-making processes. Techniques such as confidence estimation, out-of-distribution detection, conformal prediction, probabilistic inference, and causal reasoning become tightly integrated with symbolic world models, enabling systems to explicitly reason about operational risk and uncertainty rather than assuming binary correctness.

Runtime assurance must additionally support \emph{risk-aware human-robot adaptation}. Human behavior is inherently dynamic and difficult to enumerate exhaustively during design-time validation, making safe interaction dependent on continuous adaptation to human cognitive and behavioral states. Emerging embodied copilots and collaborative agents increasingly integrate multimodal interaction signals---including behavioral patterns, physiological sensing, gaze tracking, and EEG-based cognitive measurements---to estimate workload, attention, trust, intent, or cognitive stress during operation. Such mechanisms would allow for dynamic calibration of autonomy levels, interaction strategies, and safety constraints according to human state, operational risk, and context.

Another critical direction is \emph{runtime enforcement and safe fallback}. Since operational assumptions may become invalid during deployment, embodied systems must continuously ensure that safety constraints, environmental assumptions, and system requirements remain satisfied during execution. Future systems may increasingly support meta-monitoring of ``requirements of requirements''~\cite{silva2011awareness}, where systems reason not only about their behavior, but also about the validity and contextual applicability of their own operational assumptions and safety conditions. When violations, uncertainty escalation, or unsafe behaviors are detected, systems can trigger safe fallback strategies such as degraded autonomy modes, constrained operation regions, human intervention requests, or transition to conservative symbolic control policies. Runtime enforcement and fallback would thus provide an adaptive resilience layer when complete verification of open-world embodied intelligence remains infeasible.

More broadly, runtime assurance requires integrated adaptive safety architectures that continuously reason about uncertainty, human interaction, operational assumptions, and environmental context during deployment. As illustrated in Figure~\ref{fig:overview}, uncertainty-aware symbolic reasoning, risk-aware human-robot adaptation, and runtime monitoring with safe fallback together provide the foundation for runtime assurance in embodied AI systems.

\section{Toward an Integrated Research Agenda for Reliable Embodied AI}

The preceding discussions highlight that reliability in embodied AI cannot be achieved through isolated advances in testing, verification, or runtime assurance alone. Although scenario-based testing, compositional verification, and runtime assurance address complementary aspects of reliability, they are often developed independently, resulting in fragmented assurance pipelines with inconsistent assumptions, representations, and evaluation objectives. Therefore, a central challenge is to move from component-centric assurance toward \emph{integrated lifecycle assurance}. Rather than treating testing, verification, and runtime monitoring as separate activities, embodied AI systems require coherent assurance workflows spanning specification mining, scenario generation, formal reasoning, deployment monitoring, and adaptive decision-making.

A promising direction toward integrated lifecycle assurance is the development of \emph{shared neuro-symbolic representations} that connect neural perception, formal reasoning, testing coverage, and runtime monitoring across the assurance lifecycle. Structured abstractions---such as scene graphs, symbolic world models, object relations, trajectories, and domain-specific languages---can provide meaningful, interpretable interfaces across testing, verification, and deployment. Such representations support scenario-semantic coverage in testing, compositional reasoning in verification, and interpretable uncertainty-aware reasoning during deployment.

More importantly, reliable embodied AI requires \emph{cross-lifecycle feedback loops} rather than isolated assurance stages. Coverage-driven testing can expose failure modes and edge cases that are difficult to capture through formal reasoning alone, while runtime monitoring can identify assumption violations, uncertainty escalation, and anomalous behaviors during deployment. These observations can subsequently refine scenario generation, symbolic abstractions, verification assumptions, and runtime adaptation policies, enabling closed-loop assurance workflows that continuously evolve throughout the system lifecycle. Reliability in embodied AI should, thus, be viewed not as a one-time certification outcome, but as a continuously evolving assurance process driven by interactions among testing, verification, deployment monitoring, and adaptive runtime reasoning.

Beyond evaluation methodologies themselves, integrated lifecycle assurance will also require standardized assurance artifacts and reporting interfaces that improve transparency and trust across embodied AI systems. Inspired by concepts such as model cards~\cite{mitchell2019model}, structured AI test cases, and certification or rating mechanisms~\cite{weng2020towards}, future assurance workflows may need to explicitly document operational assumptions, evaluation conditions, uncertainty bounds, runtime monitoring capabilities, adaptive fallback behaviors, and known system limitations. Such artifacts can provide interpretable assurance evidence across testing, verification, and deployment stages, supporting greater transparency, reproducibility, and communication of system-level risks and guarantees.

Significant open challenges remain, as existing assurance methodologies often rely on incompatible abstractions, evaluation metrics, and assumptions about operational environments. Therefore, advancing reliable embodied AI will require interoperable symbolic representations, principled robustness metrics, shared benchmarks, standardized assurance pipelines, and closer collaboration across AI, robotics, software engineering, and formal methods communities. Progress will also depend on sustained interdisciplinary initiatives---including dedicated conference tracks, workshops, and community forums---that bring together researchers across embodied AI, testing, verification, and assurance. Ultimately, achieving reliable embodied intelligence will depend on integrated neuro-symbolic assurance ecosystems that tightly couple testing, verification, runtime monitoring, adaptive reasoning, and uncertainty-aware decision-making across the entire system lifecycle.
\clearpage

\bibliographystyle{ACM-Reference-Format}
\bibliography{acmart}

\end{document}